\documentclass[webpdf,contemporary,large]{oxford}

\usepackage{anyfontsize}
\usepackage{tikz,color,xcolor,float}
\usepackage{pgfplots}
\pgfplotsset{compat=1.7}

\usepackage{caption}
\usepackage{booktabs}
\usepackage[normalem]{ulem}

\definecolor{clr1}{RGB}{255, 172, 28}
\definecolor{clr2}{RGB}{0, 191, 255}
\definecolor{clr3}{RGB}{238, 130, 238}
\definecolor{clr4}{RGB}{32, 178, 170}
\definecolor{clr5}{RGB}{255, 99, 71} 

\graphicspath{{figures/}}

\theoremstyle{thmstyleone}%
\theoremstyle{thmstyletwo}%
\theoremstyle{thmstylethree}%

\begin{document}

\journaltitle{XXX}
\DOI{XXX}
\copyrightyear{}
\pubyear{}
\access{}
\appnotes{Original Paper}
\firstpage{1}

\title[HERMES]{Heterogeneous Entity Representation for Medicinal Synergy Prediction}

\author[1]{Jiawei Wu$^\dagger$}
\author[2]{Jun Wen$^\dagger$}
\author[1]{Mingyuan Yan$^\dagger$}
\author[3]{Anqi Dong}
\author[4]{Shuai Gao}
\author[5]{Ren Wang}
\author[6,7,*]{Can Chen}

\authormark{Wu et al.}

\address[1]{\orgdiv{School of Medicine}, \orgname{National University of Singapore, Singapore 119077}}
\address[2]{\orgdiv{Harvard Medical School}, \orgname{Harvard University, Boston, MA 02115, USA}}
\address[3]{\orgdiv{Division of Decision and Control Systems and Department of Mathematics}, \orgname{KTH Royal Institute of Technology, SE-100 44 Stockholm, Sweden}}
\address[4]{\orgdiv{Harvey Cushing Neuro-Oncology Laboratories, Department of Neurosurgery}, \orgname{Brigham and Women’s Hospital, Harvard Medical School, Boston, MA 02115, USA}}
\address[5]{\orgdiv{Department of Electrical and Computer Engineering}, \orgname{Illinois Institute of Technology, Chicago, IL 60616, USA}}
\address[6]{\orgdiv{School of Data Science and Society}, \orgname{University of North Carolina at Chapel Hill, Chapel Hill, NC 27599, USA}}
\address[7]{\orgdiv{Department of Mathematics}, \orgname{University of North Carolina at Chapel Hill, Chapel Hill, NC 27599, USA}}

\corresp[$\dagger$]{Equal contributions.}

\corresp[$\ast$]{Corresponding author.}

\received{Date}{0}{Year}
\revised{Date}{0}{Year}
\accepted{Date}{0}{Year}

\abstract{
\textbf{Motivation:} Forecasting the synergistic effects of drug combinations facilitates drug discovery and development, especially regarding cancer therapeutics. While numerous computational methods have emerged,  most of them fall short in fully modeling the relationships among clinical entities including drugs, cell lines, and diseases, which hampers their ability to generalize to drug combinations involving unseen drugs. These relationships are complex and multidimensional, requiring sophisticated modeling to capture nuanced interplay that can significantly influence therapeutic efficacy.\\
\textbf{Results:} We present a novel deep hypergraph learning method named Heterogeneous Entity Representation for MEdicinal Synergy prediction (HERMES) to predict the synergistic effects of anti-cancer drugs. Heterogeneous data sources, including drug chemical structures, gene expression profiles, and disease clinical semantics, are integrated into hypergraph neural networks equipped with a gated residual mechanism to enhance high-order relationship modeling. HERMES demonstrates state-of-the-art performance on two benchmark datasets, significantly outperforming existing methods in predicting the synergistic effects of drug combinations, particularly in cases involving unseen drugs. \\
\textbf{Availability:} The source code is publicly available at \href{https://github.com/Christina327/HERMES}{https://github.com/Christina327/HERMES}.\\
\textbf{Contact:} \href{canc@unc.edu}{canc@unc.edu}\\
}

\maketitle

\section{Introduction}
Exploring drug combinations leads a promising avenue for enhancing cancer treatment efficacy while minimizing toxicity and adverse reactions in modern medicine \citep{jia2009mechanisms, csermely2013structure,wang2012exploring,foucquier2015analysis}. Combination therapies, involving multiple drugs administered as a single treatment regimen, offer potential benefits over traditional single-drug approaches, particularly under cancer and tumor treatment contexts \citep{chou2006theoretical, o2016unbiased}. Not only do they hold the promise of greater therapeutic efficacy, but present an opportunity to mitigate host toxicity and unwanted side effects, as the doses of drug combinations are often sub-mutagenic compared to individual drug doses. However, optimizing drug combinations can be challenging, as poorly chosen combinations may lead to adverse effects and sub-optimal outcomes \citep{hecht2009randomized, tol2009chemotherapy}. Thus, there is a critical need to identify precise synergistic drug pairs tailored to different cancer types.

Historically, the identification of effective combination drugs relied on clinical experience, a process that is not only time-consuming and resource-intensive but also prone to trial and error. In contrast, high-throughput screening has emerged as an affordable and efficient strategy for identifying synergistic drug pairs, leading to the generation of extensive datasets \citep{o2016unbiased,holbeck2017national,jaaks2022effective}. However, certain limitations persist, such as the inability of cancer cell lines to accurately represent \textit{in vivo} states and the impracticality of exhaustively testing all members of the full combinatorial space with high-throughput screening \citep{ferreira2013importance,goswami2015new,morris2016systematic}.

Recently, numerous computational methods for predicting drug synergy have been proposed. Pioneering methods, including DeepSynergy \citep{preuer2018deepsynergy} and Matchmaker \citep{kuru2021matchmaker}, utilize deep neural networks with both the chemical properties of drugs and the gene expression profiles of cell lines. The deep tensor factorization model \citep{sun2020dtf} combines tensor decomposition with neural networks to forecast the synergistic effects of drug combinations. Additionally, TransSynergy \citep{liu2021transynergy} adopts a transformer network model using drug-target and protein-protein interaction data. DeepDDS \citep{wang2022deepdds} employs graph convolutional networks (GCNs) \citep{wu2020comprehensive} and multi-layer perceptrons (MLP) for synergy prediction. The current state-of-the-art method HypergraphSynergy \citep{liu2022multi} made strides in this direction by incorporating hypergraph neural networks (HGNNs) \citep{feng2019hypergraph, bai2021hypergraph} to model these dynamics in a more interconnected and multifaceted manner. Hypergraphs generalize graphs by allowing hyperedges to connect more than two nodes \citep{chen2020tensor,chen2021controllability,pickard2023hat}, encode multidimensional (or high-order) correlations and connections. 
However, these existing methods often fall short in effectively predicting drug synergy due to their neglect of higher-order drug interactions or lack of important biomedical knowledge. DeepDDS focuses solely on pairwise drug interactions using GCNs, and HypergraphSynergy, although it considers higher-order drug-drug-cell line combinations, fails to incorporate additional biomedical interactions, which hinders its generalization, particularly involving unseen drug molecules.

\begin{figure*}[b]
\centering
\includegraphics[width=\linewidth]{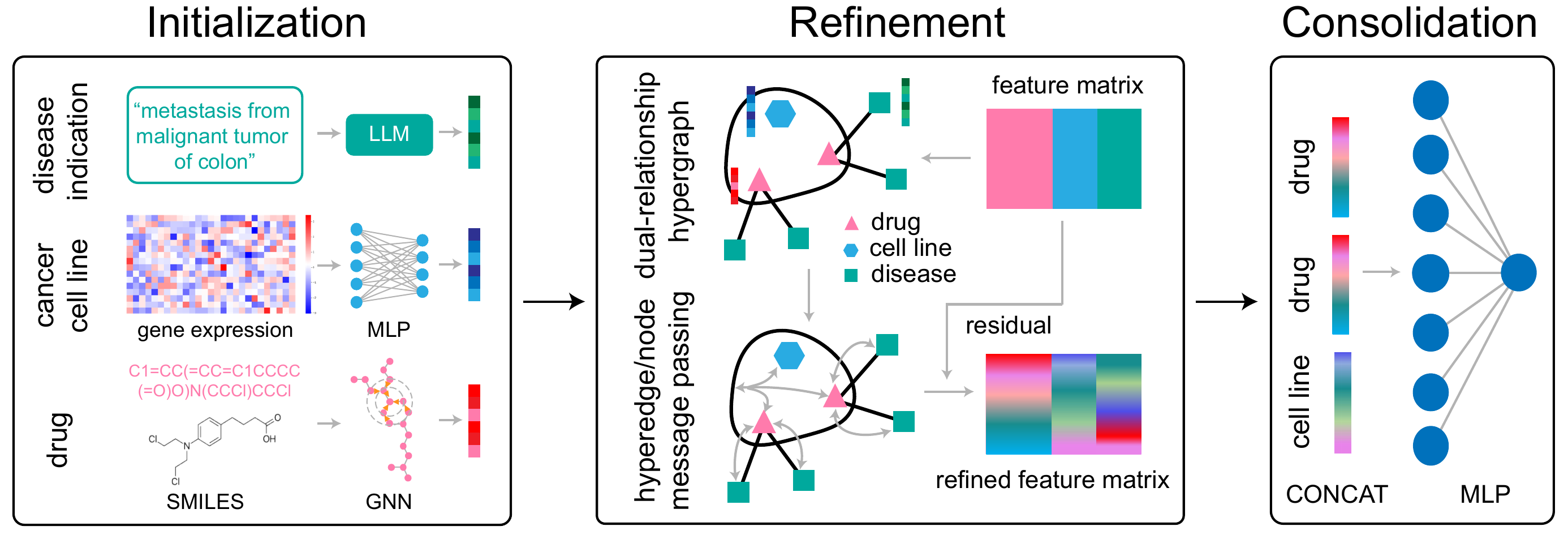}
\caption{Overview of HERMES framework. HERMES has three key phases:
(1) initialization: acquiring and transforming initial features of drugs, cell lines, and disease indications to a uniform dimensionality;
(2) refinement: enhancing feature representations through the construction of a dual-relationship hypergraph and the application of hypergraph neural networks with gated residual connections;
(3) consolidation: integrating refined features using a binary classification model to predict drug synergies with high accuracy.
Each phase is integral to the framework’s ability to process diverse data types and generate precise drug synergy predictions.}
\label{fig:drugsynergy}
\end{figure*}

In this article, we propose a novel deep hypergraph learning method – Heterogeneous Entity Representation for MEdicinal Synergy prediction (HERMES) – to enhance drug synergy prediction. HERMES distinguishes itself with its innovative strategy for integrating a variety of data sources, including drug chemical properties, cell line gene expressions, and interactions between drugs and indications. This integration is achieved through a heterogeneous hypergraph structure, which aids in assimilating extensive prior knowledge and enhances the model’s ability to generalize in novel contexts. Incorporating information about drug indications is particularly important, as some drugs may function through similar molecular mechanisms in different diseases (e.g., Bevacizumab is effective against various cancers, including colorectal cancer and non-small cell lung cancer). In contrast, drugs like glucocorticoids demonstrate different mechanisms across diseases due to distinct biological pathways, such as their anti-inflammatory action in rheumatoid arthritis versus their apoptosis-inducing role in leukemia. This approach allows for more comprehensive capturing of the varied biological contexts in which drug combinations achieve synergetic effects. Another notable breakthrough of HERMES is addressing the widespread issue of over-smoothing in message-passing networks by implementing a gated residual mechanism \citep{li2018deeper}, which not only retains more information but also notably enhances the expressiveness of the network. 

Our empirical results highlight HERMES’s effectiveness, especially in novel scenarios,  surpassing HypergraphSynergy and establishing it as a leading solution. The key contributions of this article are (1) integration of varied knowledge sources (including drugs, cancer cell lines, and disease indications) into a scalable and heterogeneous model architecture; (2) enhancement of message passing over hypergraphs with gated residual mechanisms for augmented network expressiveness; (3) empirical evidence showcasing superior performance compared to previous methods, highlighting robust generalization in novel contexts, with an average improvement of 5\% in evaluation metrics over the state-of-the-art methods.
This article is structured into four sections. The main architecture of HERMES is detailed in Section \ref{sec:2}. We assess the performance of HERMES (along with other representative drug synergy prediction methods) including a comprehensive ablation study in Section \ref{sec:3}. Finally, we conclude by discussing future research directions in Section \ref{sec:4}.


\section{Materials and Methods}\label{sec:2}
\subsection{Overview of HERMES}

Our framework conceptualizes drug synergy prediction under a hypergraph framework, treating it as a hyperedge prediction problem, where drugs, cell lines, and diseases are represented by nodes while synergistic drug–drug–cell line triplets and drug-disease pairwise relations are represented by hyperedges. Hyperedge prediction is a generalization of edge prediction on graphs \citep{chen2023survey,chen2023teasing,zhang2018beyond,sharma2021c3mm,yadati2020nhp,kumar2020hpra}. This innovative approach unfolds through three interconnected phases: initialization, refinement, and consolidation (Figure \ref{fig:drugsynergy}). Each phase plays a pivotal role in processing and integrating varied data types, thereby synthesizing comprehensive information crucial for accurate drug synergy predictions. Drawing inspiration from HypergraphSynergy, our model introduces novel methodologies and integrates cutting-edge techniques to significantly enhance prediction accuracy and reliability at each phase of analysis.

\subsection{Feature Initialization}

The initialization phase is crucial in our synergy prediction methodology, as it involves the acquisition of initial features for drugs, cell lines, and diseases. This phase performs modality-specific representation learning to transform the raw features of different modalities into a common feature space. This alignment process is critical for the subsequent hypergraph learning as it ensures consistency in the representations, allowing for effective integration of features across modalities.

\noindent \textbf{Drug features.}
We generate drug features by utilizing molecular graphs, which are built from the Simplified Molecular Input Line Entry System (SMILES) representations of drugs \citep{kim2019pubchem}. By employing the ConvMolFeaturizer method from the DeepChem library \citep{duvenaud2015convolutional}, we construct a molecular graph for each drug, with atoms represented as nodes and bonds as edges. Subsequently, we employ advanced graph transformer networks (GTNs) \citep{yun2019graph} on molecular graphs to enhance the representations of atoms and drugs. 
Denote the feature vector of atom $i$ as $\textbf{a}_i$. The updated feature (denoted as $\textbf{a}'_i$) through a GTN is  then computed as
\begin{align}
    \textbf{a}'_i = \sigma\Big(\textbf{W}_1\textbf{x}_i + \sum_{j\in\mathcal{N}(i)}\alpha_{ij} \textbf{W}_2\textbf{a}_j\Big),
\end{align}
where $\mathcal{N}(i)$ denotes the neighboring atoms of atom $i$, $\textbf{W}_1$ and $\textbf{W}_2$ are learnable weight matrices, and $\sigma$ is a nonlinear activation function. Here, $\alpha_{ij}$ is the multi-head attention coefficient defined as
\begin{equation*}
\alpha_{ij} = \text{Softmax} \Big(
\frac{(\textbf{W}_3\textbf{a}_i)^\top(\textbf{W}_4\textbf{a}_j)}{\sqrt{d}}
\Big),
\end{equation*}
where $d$ is the latent size of each head, and $\textbf{W}_3$ and $\textbf{W}_4$ are learnable weight matrices. GTNs can flexibly capture the intricate and long-range dependencies among drugs' atoms by adaptively learning attention weights among them, offering more comprehensive drug representations over traditional GCNs which are often limited by local receptive fields. Finally, we compute the feature vector of a drug by aggregating all its updated atom features using maximum pooling. The choice of maximum pooling is motivated by the fact that drug molecular graphs may vary in the number of atom nodes. The resulting feature matrix for all drugs is denoted as $\textbf{X}_{\text{drug}}$.

\noindent
\textbf{Cell line features.}
We utilize gene expression profiles to generate cell line features. Following a log2 transformation and z-score normalization of gene expression data, we employ an MLP to derive the features for cell lines, ensuring uniform dimensionality with drug features. Denote the feature vector for cell line $i$ by $\textbf{c}_i$, the updated feature (denoted as $\textbf{c}'_i$) through an MLP is computed as
\begin{equation}
    \textbf{c}'_i = \sigma(\textbf{W}_5\textbf{c}_i+\textbf{b}_5),
\end{equation}
where $\textbf{W}_5$ and $\textbf{b}_5$ are learnable weight matrix and a bias vector, respectively. The resulting feature matrix for all cell lines is denoted as $\textbf{X}_{\text{cell}}$.

\noindent
\textbf{Disease features.}
We create disease features by leveraging a novel embedding method called CODER, a knowledge-graph-guided large language model for cross-lingual medical term representation using contrastive learning \citep{yuan2022coder}. CODER excels in providing close vector representations for various terms representing similar medical concepts in multiple languages, making it particularly beneficial for extracting enriched and contextually relevant disease features \citep{yuan2022coder}. Similar to cell line features processing, we employ an MLP on the obtained disease features to align with the consistent dimensionality of drug and cell line features. The resulting feature matrix for all diseases is denoted as $\textbf{X}_{\text{dis}}$. Notably, integrating CODER with disease information enhances our model's ability to comprehend the intricate relationships between drugs, cell lines, and diseases within the hypergraph framework.

\subsection{Feature Refinement}
In the refinement phase, we focus on transforming the obtained drug, cell line, and disease features into contextually enriched representations. This process is achieved through the construction of a dual-relation hypergraph and the implementation of an advanced feature enhancement technique, namely HGNNs with gated residual connections \citep{li2018deeper}. These elements collectively elevate our model's ability to identify complex interrelationships, setting the stage for more precise synergy predictions.

\noindent
\textbf{Hypergraph construction.}
We construct a novel dual-relationship hypergraph involving relationships between drugs, cell lines, and diseases. The hypergraph contains two types of hyperedges: (1) drug-drug-cell line triplets (third-order interactions), which capture the potential synergistic effects between specific drugs and cell lines; (2) drug-disease relationships (pairwise interactions or traditional edges), which represent associations between drugs and their corresponding disease indications. In this context, diseases refer specifically to drug indications rather than the broader cancer types represented by cell lines. Therefore, we do not explicitly construct hyperedges between cell lines and diseases. This dual-relationship hypergraph, which has never been considered in previous methods, is pivotal in enriching the feature representations by integrating the various features of drugs, cell lines, and diseases.

\noindent
\textbf{Hypergraph neural networks.}
We employ advanced HGNNs with gated residual mechanisms to refine drug, cell lines, and disease features. HGNNs excel in capturing the complex relationships and interactions within the hypergraph, making it a pivotal component in our model for drug synergy prediction \citep{feng2019hypergraph}. Denote  the feature matrix as $\textbf{X}=\text{vercat}(\textbf{X}_{\text{drug}}, \textbf{X}_{\text{cell}}, \textbf{X}_{\text{dis}})$ where vercat is the vertical concatenation operation. The refined features (denoted as $\textbf{X}'$) through an HGNN are computed as
\begin{equation}
\textbf{X}'=\sigma(\textbf{D}^{-1}\textbf{H}\textbf{E}^{-1}\textbf{H}^\top\textbf{X}\textbf{W}_6),
\end{equation}
where $\textbf{D}$ and $\textbf{E}$ are diagonal matrices of node and hyperedge degrees, respectively. $\textbf{H}$ is the incidence matrix of the hypergraph, containing binary values to indicate the presence or absence of any node in any hyperedge, and $\textbf{W}_6$ is the learnable weight matrix. Here, the superscripts ${-1}$ and $\top$ denote matrix inversion and transposition, respectively. To enhance the expressive capacity of our network and extract more intricate patterns from the hypergraph, we increase the number of layers in the HGNN. However, this approach leads to a well-known challenge in graph neural networks, i.e., over-smoothing \citep{chen2020measuring}.

The phenomenon of over-smoothing occurs when the network layers become too deep, leading to the homogenization of node features and a loss of valuable information. To mitigate this issue and further empower our network's expressive capabilities, we introduce a novel solution using gated residual connections \citep{li2018deeper}. The implementation of gated residual connections within an HGNN is defined as
\begin{equation}
    \textbf{X}' = \textbf{X} + \sigma\Big(\textbf{W}_7 \sigma(\textbf{D}^{-1}\textbf{H}\textbf{E}^{-1}\textbf{H}^\top\textbf{X}\textbf{W}_6) + \textbf{b}_7\Big) \textbf{X},
\end{equation}
where $\textbf{W}_7$ and $\textbf{b}_7$ are learnable weight matrix and bias vector of the gate, respectively. The introduction of the gated residual connection in HGNNs allows our neural network to dynamically regulate the integration of original and convoluted features, which significantly reduces the risk of over-smoothing by providing a controlled blending of features. Additionally, we implement the equilibrium bias initialization  (EBI) strategy \citep{wang2010equilibrium}, where we initialize $\textbf{b}_7$ with a relatively negative value. The EBI strategy ensures that the gate functions close to an identity operation at the start of training, allowing for gradual and more effective feature integration. Our refinement approach utilizing HGNNs coupled with gated residual connections significantly enhances the model’s ability to discern complex patterns and interactions between drugs, cell lines, and diseases, thereby creating more accurate representations for them compared to previous methods such as HypergraphSynergy.

\subsection{Feature Consolidation}

In the consolidation phase, our primary objective is to produce predictive insights by integrating the refined features obtained from the earlier phase. We employ a binary classification model that uses the features refined over the dual-relationship hypergraph to identify drug synergies. It classifies the interactions between drug combinations and cell lines into two categories: synergistic and non-synergistic, which provides a clear, binary output for each potential drug synergy scenario.

\vspace{0.1in}
\noindent
\textbf{Predictive model.}
The essence of the consolidation phase is using a predictive model that leverages the features obtained in the refinement phase to evaluate the potential synergistic effects of drug combinations on specific cell lines and their implications for disease treatment. Suppose the final refined features of drug $i$, drug $j$, and cell line $k$ are denoted by $\tilde{\textbf{d}}_i$, $\tilde{\textbf{d}}_j$, and $\tilde{\textbf{c}}_k$, respectively. The synergy score (denoted as $S$) can be computed using an MLP, i.e., 
\begin{equation}
    S = \sigma\Big(\textbf{W}_8 \text{vercat}(\tilde{\textbf{d}}_i, \tilde{\textbf{d}}_j, \tilde{\textbf{c}}_k)+\textbf{b}_8\Big),
\end{equation}
where $\textbf{W}_8$ and $\textbf{b}_8$ are a learnable weight matrix and a bias vector, respectively.

To address the unordered nature of drug pairs, a critical aspect in synergy prediction, we implement a pairwise symmetric permutation augmentation strategy \citep{zhou2024permutation}. This approach involves presenting both permutations of each drug pair in our dataset. For example, if a data sample includes the combination (drug1, drug2, cell line), we also introduce (drug2, drug1, cell line) as a distinct sample. This augmentation is essential to ensure that our model is agnostic to the order of drugs, enhancing its capability to uniformly recognize synergy.

\vspace{0.1in}
\noindent
\textbf{Model training.}
Given the binary nature of our prediction task, we utilize the cross-entropy loss function \citep{mao2023cross}, a standard and effective choice for binary classification models defined as 
\begin{equation}
    \text{Loss} = -\frac{1}{N} \sum_{i=1}^{N} [y_i \log(\hat{y}_i) + (1 - y_i) \log(1 - \hat{y}_i)],
\end{equation}
where $y_i$ and $\hat{y}_i$ represent the actual and predicted label (synergistic or non-synergistic) of the $i$th sample, respectively, and $N$ is the total number of samples. The model is trained on a dataset consisting of drug pairs, cell lines, and their corresponding synergy labels. The training process involves adjusting the weights of the neural network to minimize the cross-entropy loss,  enhancing the model's ability to accurately classify drug synergies.

\subsection{HERMES and HypergraphSynergy}
Although HERMES is inspired by HypergraphSynergy, it significantly diverges from HypergraphSynergy in the following aspects:
\begin{itemize}
    \item HERMES integrates an additional knowledge source of diseases and constructs a dual-relationship hypergraph.
    \item HERMES leverages advanced GTNs to obtain drug features, while HypergraphSynergy only utilizes traditional GCNs. 
    \item HERMES improves HGNNs during the refinement phase by incorporating gated residual mechanisms to address the issue of over-smoothing. 
    \item HERMES  exploits useful learning techniques such as EBI and pairwise symmetric permutation augmentation to enhance the model's learning ability. 
\end{itemize}
To mitigate the risk of overfitting, HERMES also includes various regularization techniques, including dropout layers, weight decay,
and early stopping mechanism. Below, we show that HERMES significantly outperforms HypergraphSynergy in two drug synergy datasets, especially under the context of predicting new drug combinations.

\section{Experiments}\label{sec:3}

\subsection{Datasets}
We collected four categories of data, including drug synergy data, molecular information for drugs, genomic characteristics of cancer cell lines, and indications for drug therapy in diseases, from multiple publicly available databases. The details of these data are listed as follows:
\begin{itemize}
    \item \textbf{Drug synergy datasets:}
    We gathered data on the synergy of anti-cancer drugs from two prominent large-scale tumor screening datasets - the O'Neil dataset \citep{o2016unbiased} and the NCI-ALMANAC dataset \citep{holbeck2017national}. The O'Neil dataset comprises $23,062$ samples involving 38 unique drugs and $39$ distinct human cancer cell lines. Each sample measures Loewe synergy scores for two drugs in combination with a specific cell line. The NCI-ALMANAC dataset contains $304,549$ samples, including ComboScores for $104$ FDA-approved drugs in pairings across the NCI-60 cell line panel.
    
    \item \textbf{Drug molecular structures:}
    Information on the SMILES of drugs is obtained from the PubChem database \citep{kim2019pubchem}.
    
    \item \textbf {Gene expression in cancer cell lines:}  
    Data on gene expression in cancer cell lines are sourced from the Cell Lines Project within the COSMIC database \citep{forbes2015cosmic}. In this context, we specifically considered the expression values of $651$ genes related to the COSMIC cancer gene census. These expression values are subjected to logarithmic (log2) transformation and z-score normalization. We also utilized the CCLE \citep{barretina2012cancer} and GDSC \citep{yang2012genomics} databases as alternative data sources (see details in Supplementary Data).

\item \textbf{Drug indications:} The drug-indication annotations are sourced from PrimeKG \citep{chandak2023building}, a multimodal knowledge graph designed for precision medicine. PrimeKG integrates information from various high-quality biomedical resources, such as DisGeNET \cite{pinero2020disgenet} and DrugBank \cite{wishart2018drugbank}, as well as data from different biological scales, including disease pathways and phenotypes. It provides detailed relationships between drugs and diseases, covering indications, contraindications, and off-label uses. This comprehensive knowledge base is particularly valuable for identifying potential new applications for drugs and optimizing treatment approaches based on the underlying molecular mechanisms of diseases.

\end{itemize}

We performed a comprehensive data preprocessing on two primary drug synergy datasets -- NCI-ALMANAC and O'Neil. To ensure the quality and relevance of the data, we excluded cell lines lacking gene expression information and drugs without SMILES details. Following this preprocessing phase, the NCI-ALMANAC dataset comprises $74,139$ measurement samples of ComboScores for $87$ drugs across $55$ cancer cell lines, while the O'Neil dataset encompasses $18,950$ samples of Loewe synergy scores for $38$ drugs and $39$ cancer cell lines (Table \ref{tab:1}). Subsequently, we removed drugs from the `drug-disease dataset' that are not present in the aforementioned drug synergy datasets. This led to the extraction of indications for $82$ diseases corresponding to $37$ drugs within the NCI-ALMANAC dataset and $42$ diseases corresponding to $12$ drugs within the O'Neil dataset.

\subsection{Baselines}
We conducted a comparative analysis of our approach with representative drug synergy prediction techniques. Below is a brief overview of each of the baseline methods:
\begin{itemize}
    \item \textbf{DeepSynergy \citep{preuer2018deepsynergy}:} It utilizes a three-layer feedforward neural network to predict synergy scores, incorporating gene expression as cell line features and three types of chemical descriptors as drug features.

    \item \textbf{DTF \citep{sun2020dtf}:} It extracts latent features from the drug synergy matrix through tensor factorization and employs them to train a deep neural network model for predicting drug synergy.

    \item \textbf{HypergraphSynergy \citep{liu2022multi} (the current state-of-the-art method):} It formulates synergistic drug combinations across cancer cell lines as a hypergraph. In this hypergraph, drugs and cell lines are represented by nodes, while synergistic drug–drug–cell line triplets are represented by hyperedges. It leverages the biochemical features of drugs and cell lines as node attributes. Additionally, a HGNN is employed to learn drug and cell line embeddings from the hypergraph and predict drug synergy.

    \item \textbf{NHP \citep{yadati2020nhp}:} It is a GCN-based model specifically designed for hypergraphs to capture complex, higher-order relationships among multiple nodes. NHP employs hyperlink-aware GCN layers to transform hyperedges into clique expansions, which enables the modeling of multi-way interactions among drugs and cell lines in the drug synergy prediction task.

\end{itemize}

\subsection{Experiment Setup}
In this study, we trained and validated the models independently using the NCI-ALMANAC and O’Neil datasets. For each dataset, it is initially divided into two distinct sets: a training set, accounting for $90\%$ of the total data, and a test set comprising the remaining $10\%$. The training set undergoes a rigorous five-fold cross-validation process. This process is structured in three unique partitioning strategies to ensure comprehensive evaluation:
\begin{itemize}
    \item \textbf{Random:} Samples are randomly divided, providing a baseline assessment of model performance.
    \item \textbf{CLine:} Samples are stratified by target cell line, ensuring each fold's validation set contains unique cell lines not present in its training set.
    \item \textbf{DrugComb:}  Samples are stratified based on drug combinations. Each validation set included drug combinations not seen in the training set, although individual drugs might overlap.
\end{itemize}
The test set is used for final evaluation, whereby an unbiased assessment of the model's predictive power is ensured.

\begin{table}[t]
\centering
\caption{Statistics of the two datasets.}
\begin{tabular}{llll}
\hline
Dataset     & \#Drugs & \#CLine & \#Samples \\ \hline
NCI-ALMANAC & 87     & 55          & 74139    \\
O'Neil      & 38     & 39          & 18950    \\ \hline
\end{tabular}
\label{tab:1}
\end{table}

Additionally, for the classification task, synergy scores are converted to binary outcomes. Following established protocols \citep{preuer2018deepsynergy,sun2020dtf}, a threshold of 30 is used. Scores above this threshold indicate a positive synergy, while scores below are deemed negative. Ultimately, the model’s performance is evaluated using key metrics: area under the receiver operating characteristic curve (AUROC), area under the precision-recall curve (AUPRC), and F1-score. These three metrics together provide a comprehensive view of the model’s effectiveness in terms of precision, recall, and overall balance between precision and recall.

\subsection{Hyperparameters}
We conducted a grid search for optimized hyperparameters. Hyperparameters that yielded the highest AUROC in cross-validation were chosen for subsequent test experiments. Table \ref{tab:hyperparameters} presents the range of values considered for each hyperparameter. The optimal values (in bold) are selected based on their performance during the training and validation phases.

\begin{table}[tb!]
\centering
\caption{Hyperparameter selection (selected hyperparameters are highlighted in bold).}
\label{tab:hyperparameters}
\begin{tabular}{ll}
\hline
\textbf{Hyperparameter} & \textbf{Values}                                                        \\ \hline
learning rate           & \{1e-3, 5e-4, \textbf{2e-4}, 1e-4, 5e-5, 2e-5\}                        \\
weight decay            & \{1e-1, \textbf{1e-2}, 1e-3, 1e-4\}                                    \\
attention heads         & \{2, \textbf{4}, 8\}                                                   \\
refinement layer        & \{2, \textbf{3}, 4\}                                                   \\
interaction weight      & \{0, \textbf{0.02}, 0.05, 0.1, 0.2, 0.5, 1.0\}                         \\ \hline
\end{tabular}
\end{table}

\subsection{Performance Comparison and Analysis}

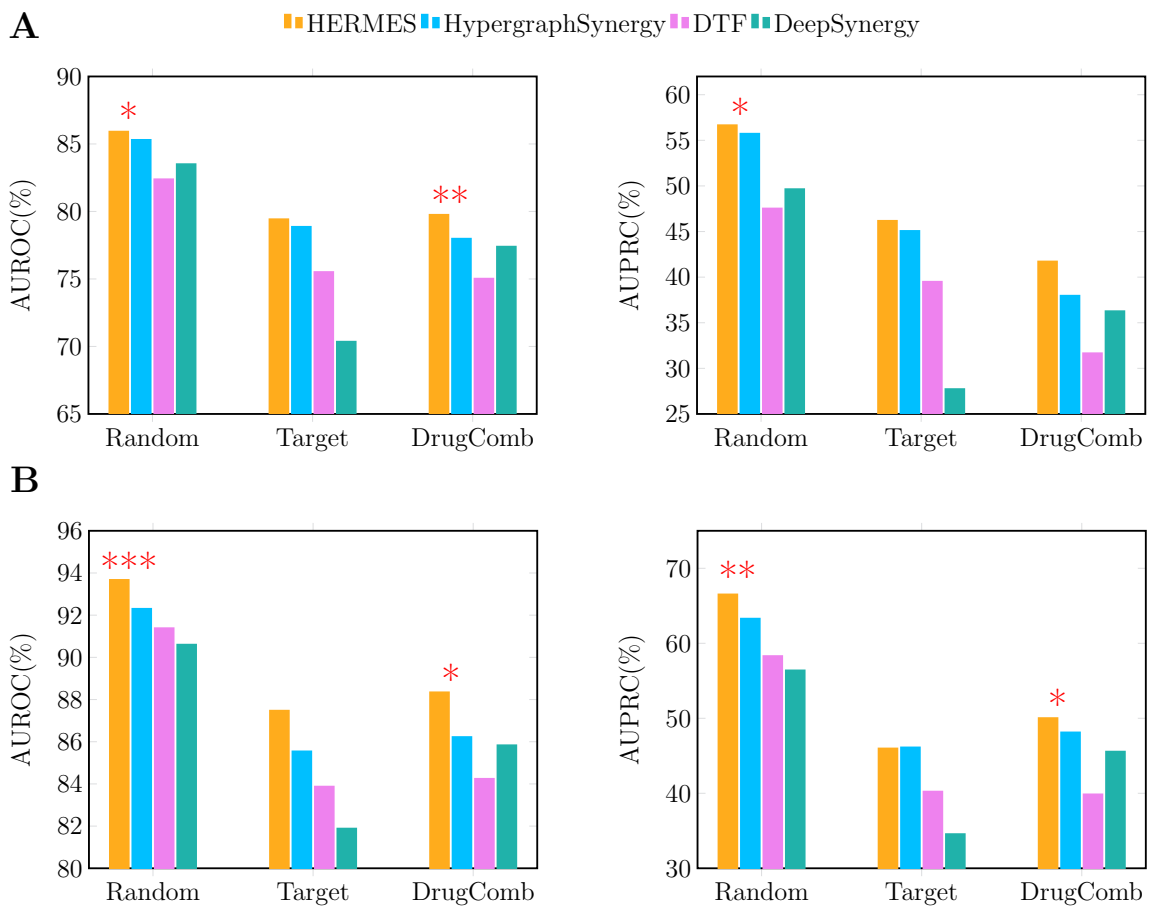
\begin{figure*}[ht!]
    \centering
      \begin{tikzpicture}[scale=0.45]
      \hspace{-2.2cm}
        \begin{axis}[
            ybar,
            bar width=10pt,
            width=12cm,
            height=8cm,
            ymin=65,
            ymax=90,
            ylabel={AUROC(\%)},
            xlabel style={font=\huge},
            ylabel style={font=\huge},
            xtick=data,
            xticklabels={Random,CLine,DrugComb},
            xticklabel style={font=\huge},
            yticklabel style={font=\huge},
            legend style={at={(1.7,1.1)},
              anchor=south,legend columns=-1,font=\huge,draw=none},
              enlarge x limits=0.25,
              line width=1pt, 
            ]    
            \addplot[fill=clr1, draw=clr1] coordinates {(1,85.91) (2,79.42) (3,79.75) }; 
            \addplot[fill=clr2, draw=clr2] coordinates {(1,85.30) (2,78.86) (3,77.98) }; 
            \addplot[fill=clr3, draw=clr3] coordinates {(1,82.38) (2,75.51) (3,75.02)}; 
            \addplot[fill=clr4, draw=clr4] coordinates {(1,83.50) (2,70.35) (3,77.39) }; 
            \addplot[fill=clr5, draw=clr5] coordinates {(1,77.69) (2,74.25) (3,75.71) };  
    \legend{HERMES, HypergraphSynergy, DTF, DeepSynergy, NHP}

            \node at (axis cs:0.85,87) [font=\huge, red] {*};
            \node at (axis cs:2.86,80.8) [font=\huge, red] {**};
        \end{axis}
        \node[font=\Large] at (-1.5,7.4) {\textbf{A}};
        \hspace{5.8cm}
        \begin{axis}[
            ybar,
            bar width=10pt,
            width=12cm,
            height=8cm,
            ymin=25,
            ymax=62,
            ylabel={AUPRC(\%)},
            xlabel style={font=\huge},
            ylabel style={font=\huge},
            xtick=data,
            xticklabels={Random,CLine,DrugComb},
            xticklabel style={font=\huge},
            yticklabel style={font=\huge},
            legend style={at={(0.5,1.1)},
              anchor=south,legend columns=-1,font=\huge},
              enlarge x limits=0.25,
              line width=1pt,
            ]
            \addplot[fill=clr1, draw=clr1] coordinates {(1,56.65) (2,46.17) (3,41.71) };
            \addplot[fill=clr2, draw=clr2] coordinates {(1,55.72) (2,45.06) (3,37.96) };
            \addplot[fill=clr3, draw=clr3] coordinates {(1,47.52) (2,39.49) (3,31.65) };
            \addplot[fill=clr4, draw=clr4] coordinates {(1,49.64) (2,27.72) (3,36.26) };
            \addplot[fill=clr5, draw=clr5] coordinates {(1,38.93) (2,35.47) (3,33.47) };         

            \node at (axis cs:0.87,58) [font=\huge, red] {*};
        \end{axis}
        \hspace{5.8cm}
        \begin{axis}[
            ybar,
            bar width=10pt,
            width=12cm,
            height=8cm,
            ymin=30,
            ymax=60,
            ylabel={F1-score(\%)},
            xlabel style={font=\huge},
            ylabel style={font=\huge},
            xtick=data,
            xticklabels={Random,CLine,DrugComb},
            xticklabel style={font=\huge},
            yticklabel style={font=\huge},
            legend style={at={(0.5,1.1)},
              anchor=south,legend columns=-1,font=\huge},
              enlarge x limits=0.25,
              line width=1pt,
            ]
            \addplot[fill=clr1, draw=clr1] coordinates {(1,53.05) (2,45.25) (3,43.43) };
            \addplot[fill=clr2, draw=clr2] coordinates {(1,52.95) (2,44.55) (3,41.28) };
            \addplot[fill=clr3, draw=clr3] coordinates {(1,46.88) (2,41.32) (3,36.13) };
            \addplot[fill=clr4, draw=clr4] coordinates {(1,48.8) (2,33.71) (3,41.04) };
            \addplot[fill=clr5, draw=clr5] coordinates {(1,42.22) (2,39.45) (3,39.21) };      

    \node at (axis cs:1.82,48) [font=\huge, red] {**};
    \node at (axis cs:2.82,45) [font=\huge, red] {***};
        \end{axis}
    \end{tikzpicture}
    
    \begin{tikzpicture}[scale=0.45]
      \hspace{-5.9cm}
        \begin{axis}[
            ybar,
            bar width=10pt,
            width=12cm,
            height=8cm,
            ymin=80,
            ymax=96,
            ylabel={AUROC(\%)},
            xlabel style={font=\huge},
            ylabel style={font=\huge},
            xtick=data,
            xticklabels={Random,CLine,DrugComb},
            xticklabel style={font=\huge},
            yticklabel style={font=\huge},
            legend style={at={(1.35,1.1)},
              anchor=south,legend columns=-1,font=\huge,draw=none},
              enlarge x limits=0.25,
              line width=1pt, 
            ]    
            \addplot[fill=clr1, draw=clr1] coordinates {(1,93.67) (2,87.47) (3,88.34) };
            \addplot[fill=clr2, draw=clr2] coordinates {(1,92.3) (2,85.54) (3,86.22) };
            \addplot[fill=clr3, draw=clr3] coordinates {(1,91.38) (2,83.87) (3,84.24)};
            \addplot[fill=clr4, draw=clr4] coordinates {(1,90.60) (2,81.88) (3,85.83) };
            \addplot[fill=clr5, draw=clr5] coordinates {(1,87.07) (2,83.42) (3,82.92) };        

            \node at (axis cs:0.85,94.3) [font=\huge, red] {***};
            \node at (axis cs:2.86,89) [font=\huge, red] {*};
        \end{axis}
        \node[font=\Large] at (-1.5,7.4) {\textbf{B}};
        \hspace{5.8cm}
        \begin{axis}[
            ybar,
            bar width=10pt,
            width=12cm,
            height=8cm,
            ymin=30,
            ymax=75,
            ylabel={AUPRC(\%)},
            xlabel style={font=\huge},
            ylabel style={font=\huge},
            xtick=data,
            xticklabels={Random,CLine,DrugComb},
            xticklabel style={font=\huge},
            yticklabel style={font=\huge},
            legend style={at={(0.5,1.1)},
              anchor=south,legend columns=-1,font=\huge},
              enlarge x limits=0.25,
              line width=1pt,
            ]
            \addplot[fill=clr1, draw=clr1] coordinates {(1,66.51) (2,45.97) (3,50.02) };
            \addplot[fill=clr2, draw=clr2] coordinates {(1,63.28) (2,46.11) (3,48.11) };
            \addplot[fill=clr3, draw=clr3] coordinates {(1,58.29) (2,40.22) (3, 39.85)};
            \addplot[fill=clr4, draw=clr4] coordinates {(1,56.38) (2,34.55) (3,45.55) }; 
            \addplot[fill=clr5, draw=clr5] coordinates {(1,47.14) (2,38.01) (3,36.30) };      
            \node at (axis cs:0.85,69) [font=\huge, red] {**};
            \node at (axis cs:2.85,52) [font=\huge, red] {*};
        \end{axis}
        \hspace{5.8cm}
        \begin{axis}[
            ybar,
            bar width=10pt,
            width=12cm,
            height=8cm,
            ymin=30,
            ymax=70,
            ylabel={F1-score(\%)},
            xlabel style={font=\huge},
            ylabel style={font=\huge},
            xtick=data,
            xticklabels={Random,CLine,DrugComb},
            xticklabel style={font=\huge},
            yticklabel style={font=\huge},
            legend style={at={(0.5,1.1)},
              anchor=south,legend columns=-1,font=\huge},
              enlarge x limits=0.25,
              line width=1pt,
            ]
            \addplot[fill=clr1, draw=clr1] coordinates {(1,62.29) (2,49.37) (3,50) };
            \addplot[fill=clr2, draw=clr2] coordinates {(1,60.25) (2,46.27) (3,48.38) };
            \addplot[fill=clr3, draw=clr3] coordinates {(1,58.28) (2,43.71) (3,43.07) };
            \addplot[fill=clr4, draw=clr4] coordinates {(1,54.45) (2,40.15) (3,47.23) };
            \addplot[fill=clr5, draw=clr5] coordinates {(1,47.68) (2,42.32) (3,40.82) };      

    \node at (axis cs:0.82,65) [font=\huge, red] {***};
    \node at (axis cs:1.82,52) [font=\huge, red] {***};
    \node at (axis cs:2.82,53) [font=\huge, red] {***};

        \end{axis}
    \end{tikzpicture}
    \vspace{0.3cm}
\caption{Model performance comparison for (A) ALMANAC Dataset and (B) ONeil Dataset. The left panels show the AUROC (\%), the middle panels present the AUPRC (\%), and the right panels display the F1-score (\%) for different models across three validation modes (Random, CLine, DrugComb). Red asterisks indicate statistical significance between HERMES and HypergraphSynergy (*** $p$-value $<0.001$; ** $p$-value $<0.01$; * $p$-value $<0.05$; two-sample $t$-test).}

\label{fig:2}
\end{figure*}

In the NCI-ALMANAC dataset, HERMES exhibits superior performance across all evaluation strategies compared to the baseline methods (Figure \ref{fig:2}A). Under the random partitioning strategy, HERMES achieves an AUROC of $85.91\%$ (std: $0.0046$), which is significantly higher than HypergraphSynergy, DTF, NHP, and DeepSynergy, which score $85.30\%$, $82.38\%$, $77.69\%$, and $83.50\%$, respectively (HERMES vs. HypergraphSynergy: $p$-value $<0.05$, two-sample $t$-test). In the more challenging `DrugComb' mode, HERMES achieves an AUROC of $79.75\%$ (std: $0.0112$), significantly outperforming the other methods, with HypergraphSynergy recording $77.98\%$ ($p$-value $<0.01$, two-sample $t$-test), NHP achieving $75.71\%$, and DTF and DeepSynergy achieving even lower scores. For the `CLine' mode, HERMES also performs better than the rest of the methods. Additionally, the AUPRC and F1-score results reflect a similar trend (middle and right panels).

The performance of HERMES is further validated using the O'NEIL dataset (Figure \ref{fig:2}B), where similar results are observed across the three modes in terms of AUROC, AUPRC, and F1-score. In the random stratification, HERMES attains an AUROC of $93.67\%$ (std: $0.0032$), significantly outperforming HypergraphSynergy, DTF, NHP, and DeepSynergy, which score $92.30\%$, $91.38\%$, $87.07\%$, and $90.60\%$, respectively (HERMES vs. HypergraphSynergy: $p$-value $<0.001$, two-sample $t$-test). In the DrugComb mode, HERMES maintains a competitive edge with an AUROC of $88.34\%$ (std: $0.0205$), significantly surpassing HypergraphSynergy at $86.22\%$ ($p$-value $<0.05$, two-sample $t$-test) and outperforming NHP at $82.92\%$. HERMES also achieves better performance under the CLine mode compared to the other methods, further supporting its robustness across different datasets.

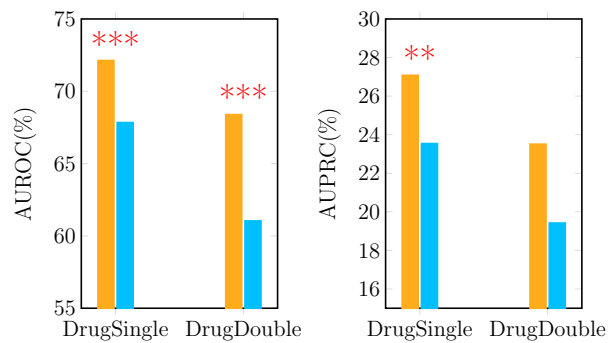
\begin{figure}
    \begin{tikzpicture}[scale=0.6]
        \begin{axis}[
            ybar,
            bar width=10pt,
            width=10cm,
            height=8cm,
            ymin=55,
            ymax=75,
            ylabel={AUROC(\%)},
            xlabel style={font=\Large},
            ylabel style={font=\Large},
            xtick=data,
            xticklabels={DrugSingle, DrugDouble},
            xticklabel style={font=\Large},
            yticklabel style={font=\Large},
            legend style={at={(0,0)},
              anchor=south,legend columns=-1,font=\Large,draw=none},
              enlarge x limits={abs=0.75cm},
              line width=1pt, 
              x=14cm
            ]    

            \addplot[fill=clr1, draw=clr1] coordinates {(1,72.13) (1.2,68.39) };
            \addplot[fill=clr2, draw=clr2] coordinates {(1,67.84) (1.2,61.04) };

            \node at (axis cs:1,73.3) [font=\huge, red] {***};
            \node at (axis cs:1.2,69.6) [font=\huge, red] {***};

        \end{axis}
        \hspace{4cm}
\begin{axis}[
            ybar,
            bar width=10pt,
            width=10cm,
            height=8cm,
            ymin=15,
            ymax=30,
            ylabel={AUPRC(\%)},
            xlabel style={font=\Large},
            ylabel style={font=\Large},
            xtick=data,
            xticklabels={DrugSingle, DrugDouble},
            xticklabel style={font=\Large},
            yticklabel style={font=\Large},
            legend style={at={(0,0)},
              anchor=south,legend columns=-1,font=\Large,draw=none},
              enlarge x limits={abs=0.75cm},
              line width=1pt, 
              x=14cm
            ]    

 \addplot[fill=clr1, draw=clr1] coordinates {(1,27.08) (1.2,23.51) };
            \addplot[fill=clr2, draw=clr2] coordinates {(1,23.54) (1.2,19.42) };

             \node at (axis cs:1,28) [font=\huge, red] {**};

        \end{axis}
    \end{tikzpicture}
    \vspace{0.3cm}
    \caption{Performance comparison of HERMES and HypergraphSynergy in the DrugSingle and DrugDouble modes using the ALMANAC dataset. Red asterisks indicate statistical significance between HERMES and HypergraphSynergy(*** $p$-value $<0.001$; ** $p$-value $<0.01$; two-sample  $t$-test).}
    \label{fig:3}
    \vspace{-15pt}
\end{figure}

To further validate the performance of our model, we considered two new validation modes using the ALMANAC dataset (Figure \ref{fig:3}) to assess the model's generalization ability to previously unseen drugs. The two modes are defined as follows: (1) \textbf{DrugSingle}, where samples are stratified based on individual drugs, with one drug in each drug combination in the validation set being novel to the training set. This approach reduces the dataset size and challenges the model to predict synergies involving new drugs; (2) \textbf{DrugDouble}, where samples are stratified such that each drug combination in the validation set contains two drugs, neither of which appears in the training set, further reducing the dataset size and increasing the difficulty. The results of `DrugSingle' and `DrugDouble' are shown in Figure \ref{fig:3}. In the `DrugSingle' mode, HERMES achieves an AUROC of $72.13\%$ (std: $0.0118$), which is significantly higher than HypergraphSynergy's $67.84\%$ ($p$-value $<0.001$, two-sample $t$-test). Similarly, in the `DrugDouble' mode, HERMES scores an AUROC of $68.39\%$ (std: $0.0468$), significantly outperforming HypergraphSynergy ($p$-value $<0.001$, two-sample $t$-test). The AUPRC metrics also show a consistent advantage for HERMES.

The comprehensive evaluation across different datasets and validation modes underscores the efficacy of HERMES in drug synergy prediction.  Although statistical insignificance is observed in certain validation modes, indicating that the extent of improvement can vary based on data distribution, sample size, and task difficulty, HERMES consistently demonstrates competitive and robust performance compared to other methods. These results validate the superiority of HERMES in addressing diverse drug synergy prediction tasks and emphasize its potential for predicting untested drug synergies in clinical applications.

\subsection{Ablation Study}
The ablation study, as detailed in Table \ref{tab:ablation}, scrutinizes the contributions of various components within the HERMES model utilizing the NCI-ALMANAC dataset. The evaluation encompasses three distinct modes: `Random', `CLine', and `DrugComb', with the primary metric being the average test AUROC for each configuration. The benchmark performance of the complete HERMES model registers an AUROC of $85.9\%$ in the `Random' mode, $79.4\%$ in the `CLine' mode, and $79.8\%$ in the `DrugComb' mode. These figures provide a baseline for evaluating the effects of systematically removing specific components or combinations of components.

\begin{table*}[t]
\caption{Ablation study results. Average test AUROC (with standard derivation) of HERMES with and without key components.}
\label{tab:ablation}
\centering
\begin{tabular}{lccc}
\toprule
Strategy & Random & CLine & DrugComb \\
\midrule
HERMES & $\textbf{85.9\%} \pm \textbf{0.005}$ & $\textbf{79.4\%} \pm \textbf{0.017}$ & $\textbf{79.8\%} \pm \textbf{0.011}$ \\
w/o transformer & $82.8\% \pm 0.007$ & $78.5\%\pm0.013$ & $78.3\%\pm 0.011$ \\
w/o disease & $85.7\%\pm 0.006$ & $79.2\%\pm 0.016$ & $79.5\%\pm 0.007$ \\
w/o gated residual & $84.0\%\pm 0.060$ & $79.0\%\pm 0.014$ & $78.8\%\pm 0.009$ \\
w/o gate & $82.5\% \pm 0.006$& $78.7\%\pm 0.013$ & $78.5\%\pm 0.009$ \\
w/o transformer + w/o disease & $82.1\%\pm 0.005$ & $77.8\%\pm 0.015$ & $77.3\%\pm 0.021$ \\
w/o transformer + w/o gate & $82.0\%\pm 0.004$ & $77.7\%\pm 0.014$ & $77.2\%\pm 0.021$ \\
w/o transformer + w/o disease + w/o gate & $80.4\%\pm 0.006$ & $76.5\%\pm 0.016$ & $77.8\%\pm 0.024$ \\

\bottomrule
\end{tabular}
\end{table*}

\noindent \textbf{Contribution of transformer architecture.} Excluding the transformer component results in significant performance drops  across all three modes. These findings highlight the indispensable role of the transformer in capturing intricate relationships within the data.

\noindent \textbf{Impact of disease knowledge}. The exclusion of disease (drug-indication relation) demonstrates a slight yet notable impact on the model's performance. These findings suggest that incorporating disease knowledge provides a performance advantage, but the margin remains relatively narrow. This may stem from the single-modal nature of the current disease data and the limitations of the classical CODER model. Exploring the integration of multimodal disease knowledge and adapting more advanced large language models could potentially enhance this performance benefit.

\noindent\textbf{Role of gated residuals.} Omitting the gated residual connections results in a notable performance degradation, particularly in the `Random' mode. This emphasizes the significance of gated residuals in maintaining high predictive accuracy, especially in scenarios with higher data variability.

\noindent \textbf{Importance of gating mechanisms.} The removal of the gating mechanism induces the most pronounced decline in performance across all modes. This underscores the critical function of gating mechanisms in modulating information flow and enhancing the model's resilience.

\noindent \textbf{Effect of weighted hyperedges.} The absence of weighted hyperedges also leads to considerable performance decline across all three modes. This underscores the necessity of weighted hyperedges for accurately modeling complex interactions, particularly in the diverse Random mode.

In summary, the ablation study clearly demonstrates that each component of the HERMES model contributes uniquely to its overall efficacy. Transformer architecture, molecule indication relation, gated residuals, gating mechanisms, and weighted hyperedges collectively enhance the model's predictive performance. These results affirm the sophisticated design of HERMES, showcasing the synergistic effect of its components in achieving superior predictive accuracy across various validation strategies.

\section{Conclusion}\label{sec:4}

In this article, we introduced HERMES, a novel deep hypergraph learning framework for drug synergy prediction that integrates heterogeneous biomedical data, including drug molecular structures, gene expression profiles of cell lines, and disease indications. Our results demonstrate that HERMES consistently achieves superior performance across two large-scale benchmark datasets, excelling particularly in challenging scenarios with previously untested drug combinations. The model’s gated residual mechanism mitigates over-smoothing in message-passing, enabling it to capture intricate high-order relationships among drugs, cell lines, and diseases with enhanced precision.

While HERMES establishes a flexible and general framework for drug synergy prediction, there are areas for improvement. First, the model currently incorporates only drug, cell line, and disease information, limiting its ability to leverage other relevant biomedical knowledge, such as protein–protein interactions, drug–target interactions, and disease–gene interactions. Integrating such information would further enrich the hypergraph structure and improve the model’s generalizability and predictive accuracy. Additionally, HERMES’s computational demands, particularly in terms of CPU and GPU usage, could present challenges when scaling to extremely large datasets or incorporating more data sources. To address this, innovative graph/hypergraph sampling techniques and optimized memory management strategies should be explored. Finally, HERMES presently models interactions between only two drugs. Extending the framework to accommodate multi-drug combinations could open promising new avenues for exploring complex treatment regimens and identifying synergies in multi-drug therapies.

In summary, HERMES presents a scalable and adaptable approach to drug synergy prediction, with potential applications extending into other biomedical and clinical research domains, such as drug discovery, personalized medicine, and the development of more effective and efficient therapeutic strategies. It enhances the ease of laboratory implementation while also providing cost-effective solutions for assessing drug synergy. The framework’s flexibility opens further opportunities to incorporate additional data sources and adapt to diverse predictive tasks, making it a foundational step towards more accurate and generalizable drug synergy models.

\bibliographystyle{apalike}
\bibliography{references}

\end{document}